\documentclass[a4paper,11pt]{article}
\usepackage{pos}

\title{The chiral phase transition at non-zero imaginary
baryon chemical potential for different numbers
of quark flavours}
\ShortTitle{Chiral phase transition for imaginary baryon chemical potential}

\author*[a,b]{Alfredo D'Ambrosio}
\author[a,b]{Reinhold Kaiser}
\author[a,b]{Owe Philipsen}

\affiliation[a]{Institut für Theoretische Physik - Goethe-Universität, \\ Max-von-Laue-Str. 1, 60438 Frankfurt am Main, Germany }
\affiliation[b]{ John von Neumann Institute for Computing (NIC) GSI, \\ Planckstr. 1, 64291 Darmstadt, Germany}

\emailAdd{ambrosio@itp.uni-frankfurt.de}

\abstract{The so-called Columbia plot summarises the order of the QCD thermal transition as a function of the number of quark flavours and their masses. Recently, it was demonstrated that the first-order chiral transition region, as seen for $N_f \in [ 3,6 ]$ on coarse lattices, exhibits tricritical scaling while extrapolating to zero on sufficiently fine lattices. Here we extend these studies to imaginary baryon chemical potential. A similar shrinking of the first-order region is observed with decreasing lattice spacing, which again appears compatible with a tricritical extrapolation to zero.}

\FullConference{%
The 39th International Symposium on Lattice Field Theory,\\
8th-13th August, 2022,\\
Rheinische Friedrich-Wilhelms-Universität Bonn, Bonn, Germany
}

\begin{document}
\maketitle

\section{Introduction}

\begin{figure}[t]
    \centering
    \includegraphics[width=1\textwidth]{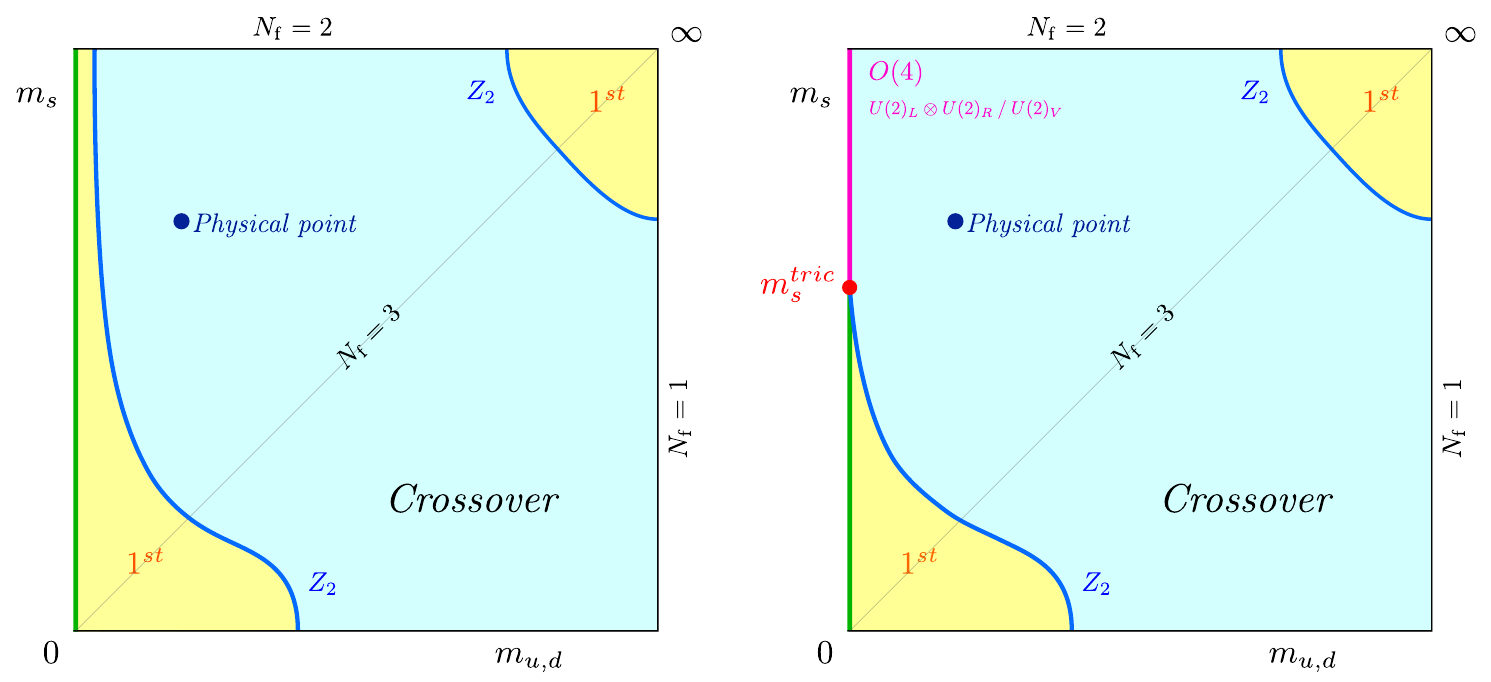}
\caption{Two plausible scenarios for the Columbia plot proposed in \cite{1}. From \cite{7}. }
\label{one}
\end{figure}
 
The order of the QCD chiral phase transition in the chiral limit has been a challenging topic for the last decades. Since the mid-eighties two plausible scenarios have been predicted for $N_f=2$, either a first-order or a second-order phase transition, depending on the anomalous $U(1)_A$ axial symmetry, whereas a first-order phase transition was predicted for $N_f \geq 3$~\cite{1}. For $N_f = 2 + 1$, the order of the QCD thermal phase transition at zero density is shown in the Columbia plot \cite{2}, where the nature of the QCD thermal transition is depicted, with first-order regions and a crossover region, separated by 3D Ising ($Z_2$) - lines of critical masses. The ambiguity for $N_f=2$ is reflected in the left area of the plots shown in figure \ref{one}, with a possible first-order region for all strange quark masses, or the presence of a tricritical point, where the first-order region meets the second-order line in the chiral limit. Many studies using different discretisations have been devoted to decide between these scenarios, with apparently conflicting results. A general trend is that finer lattices or highly improved actions make the first-order region shrink. A detailed discussion and references can be found in \cite{3,4,5}.

\begin{figure}[t]
\begin{minipage}[t]{0.5\linewidth}
    \centering
    \includegraphics[width=1\textwidth]{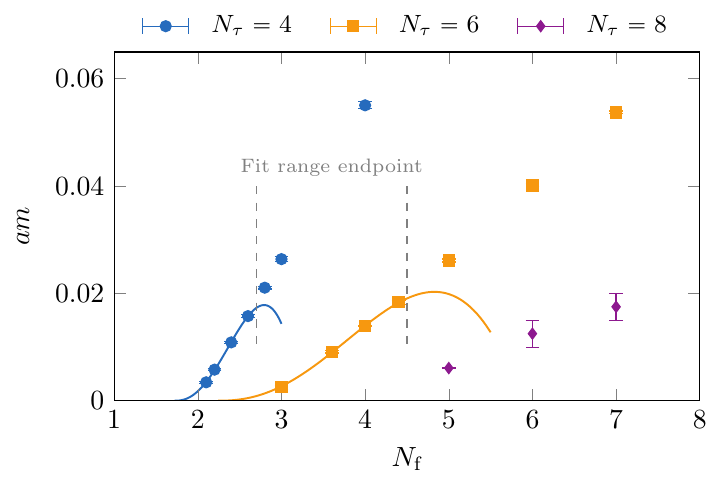}
    \caption{The $(am, N_f)$ plane with $N_f$ as a continuous parameter. The points represent the critical masses, $am_{c}$. The fitting lines correspond to the extrapolation towards the chiral limit, compatible with a tricritical scaling. From~ \cite{7}.} 
    \label{amnf}
\end{minipage}
\hspace{0.3cm}
\begin{minipage}[t]{0.5\linewidth}
    \centering
    \includegraphics[width=1\textwidth]{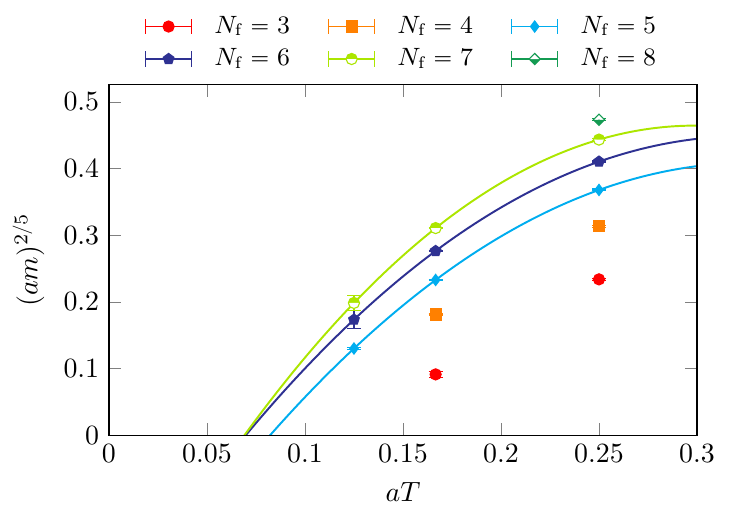}
    \caption{The critical masses  $am_{c}$ represented in the $((am)^{2/5}, aT)$ plane, compatible with a tricritical scaling. From \cite{7}.}
    \label{amnf5}
\end{minipage}
\end{figure}

QCD for $N_f$ degenerate flavours of quarks at zero density is described by the following partition function
\begin{equation}
Z(\beta,am,N_f,N_\tau) = \int \mathcal{D} U (\det D[U])^{N_f} e^{-S_g[U]},        
\end{equation}
where $\beta = 6/(g(a))^2$ is the lattice gauge coupling, $am$ are the bare quark masse on the lattice and $N_\tau= (a(\beta)T)^{-1}$ is the lattice temporal extent, $T$ being the temperature. A new approach presented in \cite{6,7}, consisting in the analytic continuation of the $N_f$ parameter from integer to continuous values, was able to resolve the question about $N_f=2,3$. The results from this work are shown in the $(am,N_f$) plane presented in figure \ref{amnf}, where the points correspond to different $Z_2$-critical masses $am_{c}$ for different $N_f$ values, and separate the crossover region above from the first-order region below, for $N_\tau \in \{4,6,8\}$. The fitting lines for $N_\tau=\{4,6\}$ follow the equation
\begin{equation}
\label{am}
am(N_\tau, N_f) \approx \mathcal{B}_1(N_\tau)(N_f -N_f^{{\mathrm{tric}}})^{5/2} + \mathcal{B}_2(N_\tau)(N_f -N_f^{{\mathrm{tric}}})^{7/2},
\end{equation}
and represent the extrapolation to the lattice chiral limit. 
Note that a tricritical point must exist whenever the chiral transition in the massless limit changes from first to second order.
The exponents reflect tricritical scaling for sufficiently small $am_{c}$ values, whose observation allows to identify and locate such a tricritical point 
${N_f}^{\mathrm{tric}}(N_\tau)$  along the x-axis. 
As shown in figure \ref{amnf5}, the same data can be depicted in the $((am)^{2/5}, aT)$ plane, with  $aT={N_\tau}^{-1}$. 
Here again the first-order region below the critical masses is separated from the crossover above,  and  the bare quark masses have been rescaled to represent the tricritical scaling field, $(am)^{2/5}$. The extrapolation towards the chiral limit is performed according to
\begin{equation}
\label{am25}
am(N_\tau, N_f)^{2/5} \approx \mathcal{A}_1(N_f)(aT -aT_{\mathrm{tric}}) + \mathcal{A}_2(N_f)(aT -aT_{\mathrm{tric}})^2,
\end{equation}
and results in $aT_{\mathrm{tric}}(N_f)$. The last result is crucial to give a resolution to the ambiguous continuum situation in figure \ref{one}. Indeed, for $N_f \in [2,6]$ the extrapolation towards the lattice chiral limit is always compatible with a finite $aT_{\mathrm{tric}}(N_f)$. On the other hand, the lines of constant physics for any  quark mass have their continuum limit in the origin  of the same plot \cite{7}. Then, also for $N_f \in [2,3]$ in the Columbia plot, the chiral phase transition must be of second order.

\begin{figure}[t]
\begin{minipage}[t]{0.40\linewidth}
    \centering
    \includegraphics[width=1\textwidth]{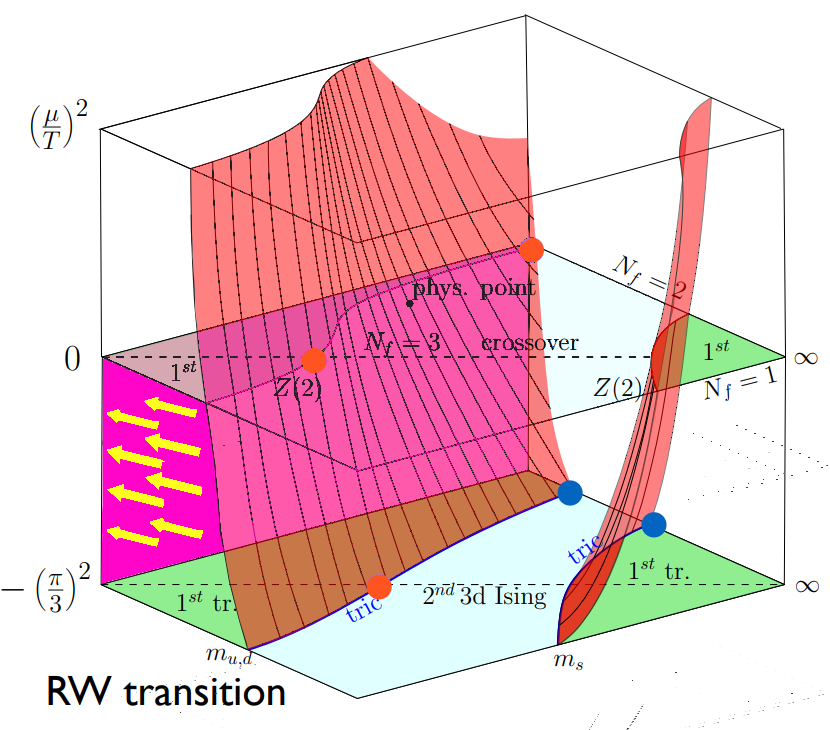}
    \caption{Schematic representation of the 3D Columbia plot for $N_\tau=4$, where negative $(\mu/T)^2$ means imaginary baryon chemical potential. For $\mu/T=0$, the 2D Columbia plot is represented, whereas for $\mu/T= i \pi/3$ it is the Roberge-Weiss plane. From \cite{9}.} 
    \label{3Done}
\end{minipage}
\hspace{0.1cm}
\begin{minipage}[t]{0.60\linewidth}
    \centering
    \includegraphics[width=1\textwidth]{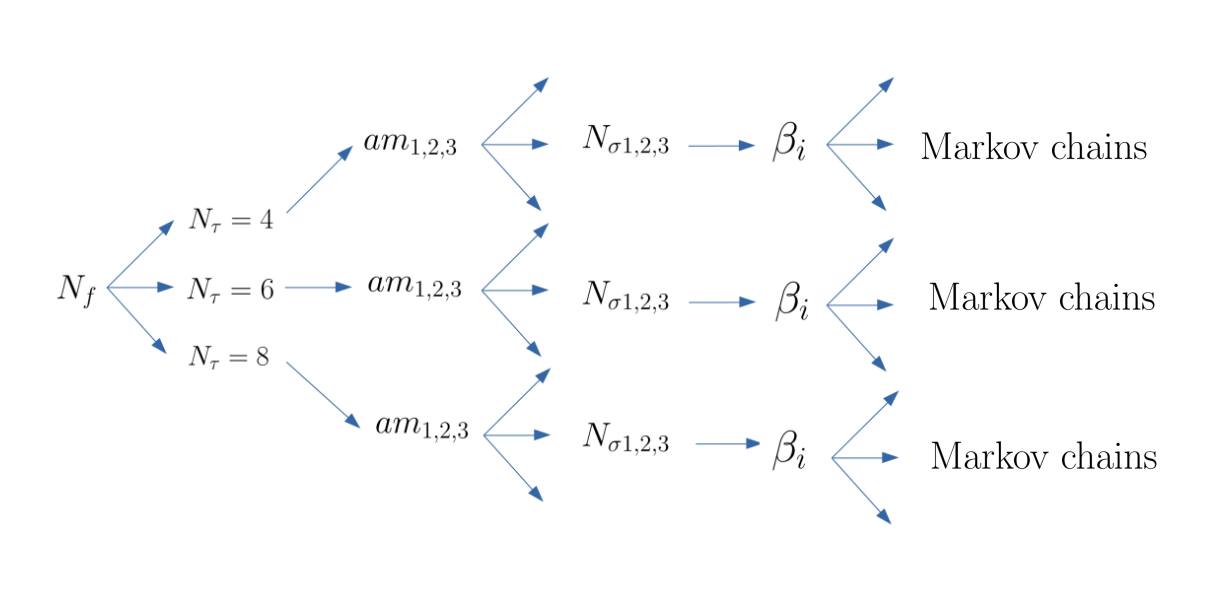}
    \caption{Sketch of the procedure used to collect data. }
    \label{flow}
\end{minipage}
\end{figure}

An extension of the Columbia plot is given through the introduction of a non-zero imaginary quark chemical potential, $\mu=i\mu_i$. The partition function obeys
\begin{equation}
Z(\mu) = Z(-\mu) \hspace{2cm} Z \left(\frac{\mu}{T}\right) = Z\left(\frac{\mu}{T} + i\frac{ 2\pi k }{3} \right),  \hspace{0.2cm}   k \in \mathbb{Z} \hspace{0.4cm}
\end{equation}
where the left equation represents the charge-parity symmetry, whereas the right one corresponds to the Roberge-Weiss periodicity \cite{8}. This provides a third axis as in figure \ref{3Done}: The $\mu=0$ surface of the cube is the usual 2D Columbia plot, whereas moving down along the $(\mu/T)^2$~-axis means to vary the value of $\mu_i$ until the bottom surface, corresponding to the Roberge-Weiss plane with $\mu_i= \pi T/3$, is reached. Both first-order regions in the Columbia plot grow while approaching the Roberge-Weiss plane on coarse $N_\tau=4$ lattices \cite{9,10,11,12,13,14,15,16,17}. 

In this work we investigate the order of the chiral phase transition as a function of lattice spacing using $N_f$ as a continuous parameter when $\mu_i = 0.81 \pi T/3$. We then study the $Z_2$ critical surface and observe how it approaches the chiral limit. At the end, we will compare the results with the ones at zero density from~\cite{7}. 

\section{Strategy}

The strategy we follow to collect data is summarised in figure \ref{flow}. The first step consists in setting $N_f \in \{2.3,3.3,3.6,4.0,4.5,5.0\}$ and for any fixed $N_f$, the lattice temporal extent is fixed such that $N_\tau \in \{4,6,8\}$. For any $N_\tau$ value, at least three bare quark masses $am$ are chosen to do a scan. At this point simulations are performed for aspect ratios $N_\sigma/N_\tau = 2,3,4$ and we scan for a set of lattice gauge coupling values, $\beta$. In our simulations, we use the unimproved staggered fermion action. 

The observable we use as approximate order parameter is the chiral condensate $\mathcal{O}=\langle \bar{\psi} \psi \rangle$, whose sampled distribution is studied through standardised moments. In general, we have 
\begin{equation}
\langle \mathcal{O} \rangle = Z^{-1} \int \mathcal{D}U  \mathcal{O}[U] (\mathrm{det}D[U])^{N_f} e^{-S_g[U]}
\end{equation}
and the standardised moments are defined by
\begin{equation}
B_n(\beta, am, N_\sigma)= \frac{\langle ( \mathcal{O} - \langle \mathcal{O}\rangle)^n \rangle}{\langle ( \mathcal{O} - \langle \mathcal{O}\rangle)^2 \rangle^{\frac{n}{2}}}.
\end{equation}
The third standardised moment is the skewness  $B_3(\beta, am, N_\sigma)$ and allows to identify the value of the (pseudo-)critical $\beta_c$ value corresponding to the phase boundary by solving the equation
\begin{equation}
B_3(\beta = \beta_c, am, N_\sigma) =0.
\end{equation}
Simulations are carried out for two to four $\beta$-values per mass of interest and volume $N_\sigma$, and Ferrenberg-Swendsen reweighting \cite{19} is employed to improve the resolution in the determination of $\beta_c$. 
The last step consists in identifying the order of the chiral phase transition corresponding to the simulated parameter setup. This is given by evaluating the fourth standardised moment, the kurtosis
\begin{equation}
B_4(\beta_c(am), am, N_\sigma).
\end{equation}
In the infinite volume limit, the kurtosis values differ discontinuously with respect to the order of the phase transition and are summarised in table \ref{tab:Table}. Simulations are always performed for finite lattice volumes $N_\sigma$, where the discontinuity between those values is 
smoothed. Taking this into account, the kurtosis evaluated at $\beta_c$ can be expanded  for large enough volumes $N_\sigma$  in the scaling variable $(am - am_{c}) N_\sigma^{1/\nu}$ \cite{20},
\begin{equation}
B_4(\beta_c, am, N_\sigma) \approx  (1.604 + c (am - am_c) N_\sigma^{1/\nu})(1+b N_\sigma ^{-y}),
\end{equation}
with a critical exponent given in table~\ref{tab:Table}.
This formula contains a subleading finite volume correction term depending on the 3D Ising critical exponent $y = 0.8940$ \cite{21}. 
\begin{table}[]
\centering
\begin{tabular}{ |c|c|c|c| } 
 \hline
  & Crossover & $1^{st}$ order & $3D$ Ising\\ 
 \hline
 $B_4$ & $3$ & $1$ & $1.604(1)$ \\ 
 $\nu$ & $-$ & $1/3$ & $0.6301(4)$\\ 
 \hline
\end{tabular}
\caption{List of kurtosis values in the infinite volume limit and critical exponent $\nu$. The $3D$ Ising values are from \cite{18}.}
\label{tab:Table}
\end{table}

Our simulations are performed using the RHMC algorithm which is implemented in our publicly available $\mathrm{CL}\textsuperscript{2}\mathrm{QCD}$ code based on OpenCL, in its latest version v1.1 \cite{22,23}. Also  multiple pseudofermions \cite{24} were involved in the simulations, in order to be able to use the RHMC algorithm also when $N_f(\mathrm{mod}4)=0$.
All of the simulations have been handled by using the software BaHaMAS \cite{25}.

\section{Results}

\begin{figure}[]
\begin{minipage}[t]{0.50\linewidth}
    \centering
    \includegraphics[width=1\textwidth]{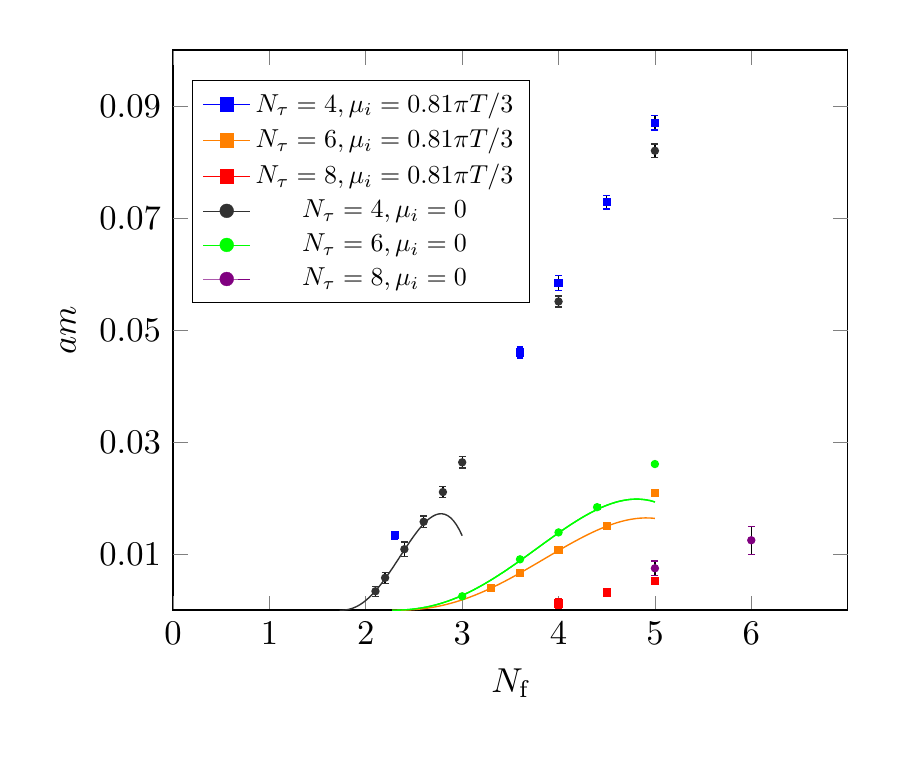}
    \caption{Critical masses separating the first \\
             order region below from the crossover above \\
             in the $(am, N_f)$ plane. Here we give a direct \\comparison 
             between the $\mu_i = 0.81\pi T/3$ and \\$\mu_i=0$ results.} 
    \label{three}
\end{minipage}
\begin{minipage}[t]{0.49\linewidth}
    \centering
    \includegraphics[width=1\textwidth]{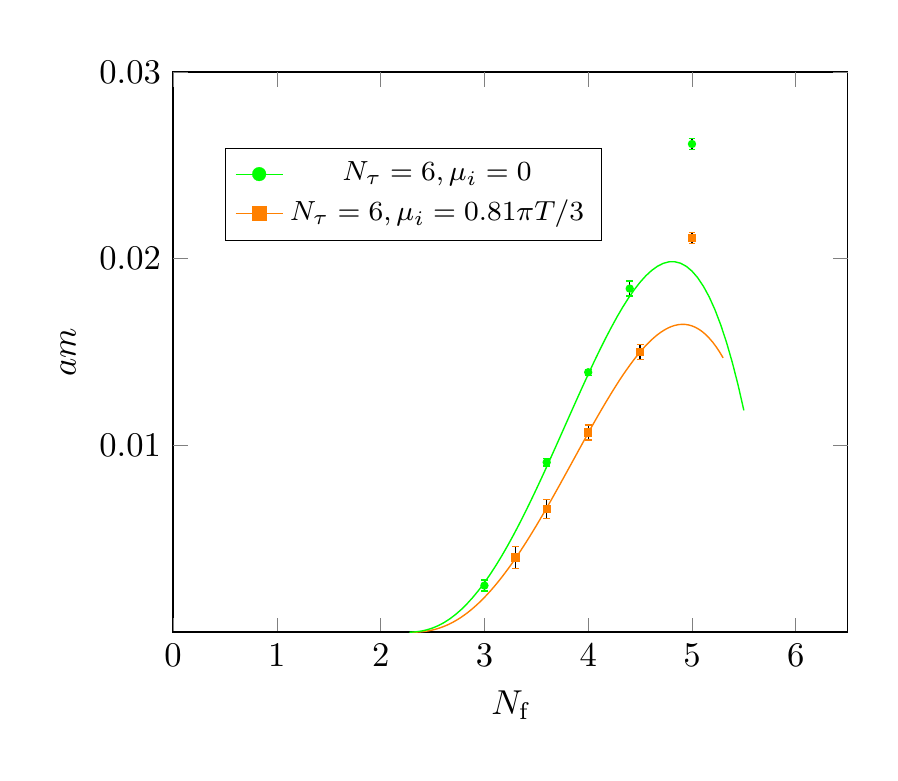}
    \caption{Detailed comparison between the results from $\mu_i = 0.81\pi T/3$ and $\mu_i=0$ in the $(am, N_f)$ plane for $N_\tau=6$.}
    \label{four}
\end{minipage}
\end{figure}

The preliminary results coming from our simulations are summarised in figure \ref{three}. For $\mu_i \neq 0$ the data for $N_\tau=6$ allow to perform the first tricritical extrapolation, according to 
\begin{equation}
am(N_\tau, N_f,\mu_i) \approx \mathcal{B}_1(N_\tau, \mu_i)(N_f -N_f^{{\mathrm{tric}}})^{5/2} + \mathcal{B}_2(N_\tau,\mu_i)(N_f -N_f^{{\mathrm{tric}}})^{7/2},
\end{equation}
which is the analogue of equation \eqref{am}, but the coefficients now depend on $\mu_i$. For $N_\tau=8$ we might already be in the tricritical scaling region, but more points and statistics are needed to draw conclusions. For $N_\tau=4$, the tricritical scaling region is expected to be detectable in the proximity of $N_f=2$.

From figure \ref{3Done} one should expect a larger first-order region when moving down along the $(\mu/T)^2$ axis for $N_\tau=4$ \cite{9}. Indeed, comparing the results in figure \ref{three} with the ones coming from $\mu_i =0$ simulations, we observe for $N_\tau=4$  the critical $Z_2$-masses  to be larger than the ones resulting from zero density.  However, for $N_\tau =6,8$ the  opposite  applies: this highlights the mixed dependence of the size of the first-order region on both 
the lattice cutoff $a$ and $\mu_i$. 
Nevertheless, a similar tricritical scaling is observed when extrapolating to the chiral limit for non-zero density, and it is shown in detail for $N_\tau=6$ in figure \ref{four}.

\begin{figure}[]
\begin{minipage}[t]{0.49\linewidth}
    \centering
    \includegraphics[width=1\textwidth]{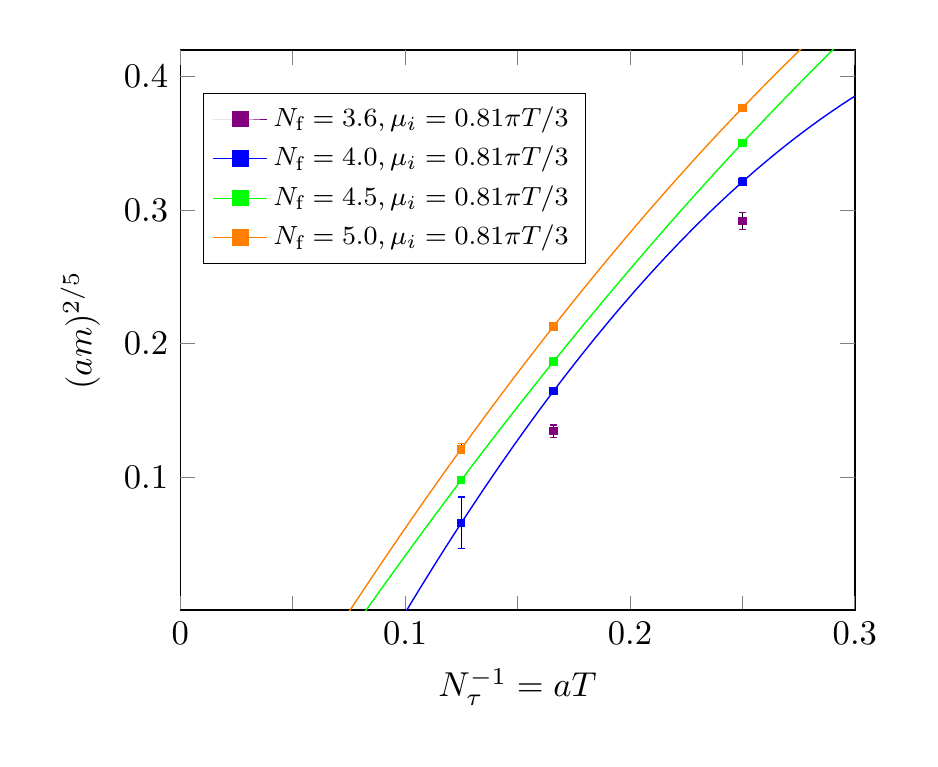}
    \caption{The scaled critical masses as functions of the lattice spacing for $N_f= 3.6,4.0,4.5,5.0$ and $\mu_i = 0.81 \pi T/3$.} 
    \label{five}
\end{minipage}
\hspace{0.1cm}
\begin{minipage}[t]{0.49\linewidth}
    \centering
    \includegraphics[width=1\textwidth]{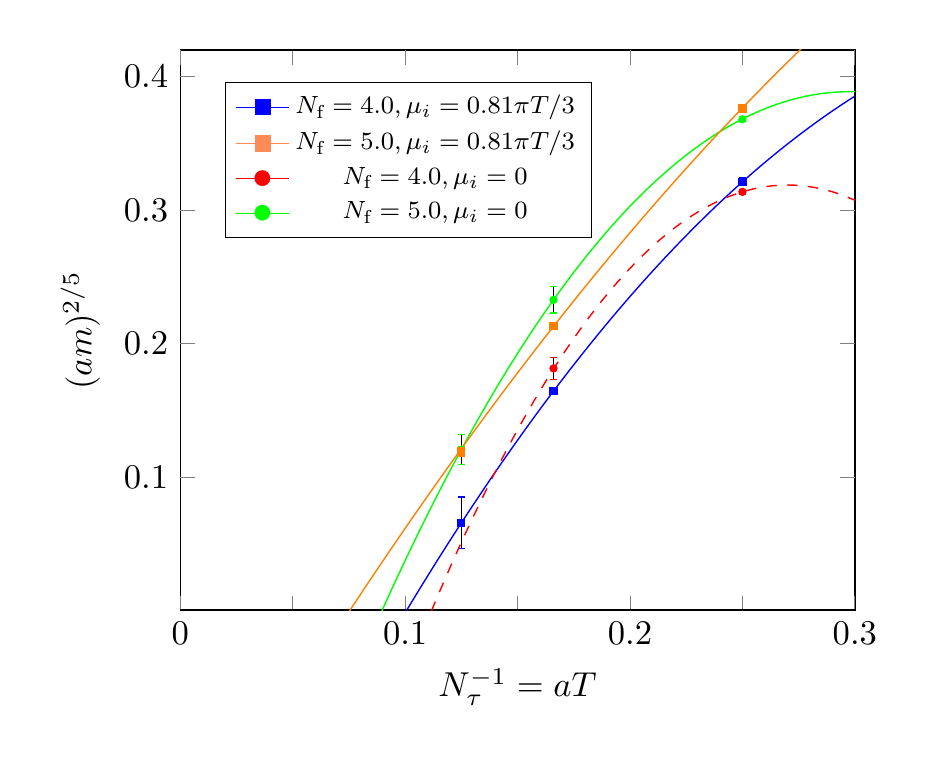}
    \caption{Comparison of the results from $\mu_i=0$ and $\mu_i = 0.81 \pi T/3$ for $N_f=4.0, 5.0$. The dashed line for $N_f=4.0$ and $\mu_i=0$ is to guide the eye }
    \label{six}
\end{minipage}
\end{figure}

A translation of these data can be provided in the $((am)^{2/5}, aT)$ plane in figure \ref{five}. In this case the critical masses along the y-axis are scaled to represent the tricritical scaling field according to
\begin{equation}
\label{eqn:25}
am(N_\tau, N_f,\mu_i)^{2/5} \approx \mathcal{A}_1(N_f, \mu_i)(aT -aT_{\mathrm{tric}}) + \mathcal{A}_2(N_f,\mu_i)(aT -aT_{\mathrm{tric}})^2,
\end{equation}
in analogy to equation \eqref{am25}. This clearly suggests the presence of a finite $aT_\mathrm{tric}(N_f)$ along the x-axis, which implies the absence of first-order regions in the continuum limit, as for $\mu_i=0$ in figure~\ref{amnf5}. A comparison between zero and finite $\mu_i$ is given in figure~\ref{six}, for $N_f=4.0$ and $N_f=5.0$. The trend for both cases is the same, except for a difference in the shape of the fitting functions, due to the dependence of equation \eqref{eqn:25} on the  parameter $\mu_i$. A similar observation has been made in the Roberge-Weiss plane \cite{26}.

\section{Conclusions}

We have shown first results on the fate of the first-order chiral transition region,
which is observed for unimproved staggered fermions with imaginary chemical
potential on coarse lattices, as a function of 
the number of degenerate quark flavours and the lattice spacing.
In complete analogy to a previous study at zero density~\cite{7}, we observe a strengthening of
the chiral transition with increasing number of quark flavours and with increasing lattice 
spacing. Conversely, our preliminary data are fully consistent with the chiral 
first-order transition terminating at some tricritical $N_\tau^\mathrm{tric}(N_f)$,
which implies that the corresponding first-order transition is a cutoff effect and not
connected to the continuum limit, just as at zero density. In the extended Columbia plot 
in figure~\ref{3Done}, the entire critical surface, at least for zero and imaginary chemical
potential, appears to move towards the zero
mass limit as the continuum is approached, as indicated by the arrows.
Note that this is also fully consistent with recent simulations of improved staggered fermions 
in the Roberge-Weiss
plane, where no first-order transition could be seen at all for $m_\pi> 40$ MeV \cite{27,28}.

\section{Acknowledgements}

We thank the staff of the VIRGO cluster at GSI Darmstadt, where all simulations have been performed. The authors acknowledge support by the Deutsche Forschungsgemeinschaft (DFG, German Research Foundation) through the CRC-TR 211 'Strong-interaction matter under extreme conditions'– project number 315477589 – TRR 211, and
by the Helmholtz Graduate School for Hadron and Ion Research (HGS-HIRe) .

\end{document}